\documentclass{elsart}
\usepackage{graphicx}

\def\bibcode#1{(\texttt{#1})}

\def\astrobj#1{#1}

\begin{document}
\begin{frontmatter}
\title{On the efficiency of the Ultra Steep Spectrum tecnique 
in finding High-z Radiogalaxies}

\author{Pedani, M.\thanksref{email}}

\address{INAF-Centro Galileo Galilei, Po Box 565, S/C La Palma - 38700 
TF, Spain}
\thanks[email]{E-mail: pedani@tng.iac.es}

\begin{abstract}
In the last three decades, the Ultra Steep spectrum tecnique has been exploited by 
many groups since it was demonstrated that radio sources with very steep spectra 
($\alpha < -1.0$; $S \propto \nu^{\alpha}$) are good tracers of high-z radio galaxies 
(HzRGs; $z > 2$). 
Though more than $150$ HzRGs have been discovered up to now with 
this tecnique, little is known about its real effectiveness, as 
most of the ongoing searches still have incomplete follow-up programs. 
By selecting a new appropriate sample of USS sources from the MRC survey, the true 
searching efficiency of the USS tecnique has been quantitatively demonstrated for 
the first time in this paper. 
Moreover it was compared with that of an optical search of HzRGs 
based on a simple cut of the galaxies r-band magnitude distribution. 
When no bias other than the radio-spectrum steepness is applied, the USS 
tecnique may be up to $4$ times more efficient in selecting HzRGs with respect to an 
optical search. 
Nevertheless, when the search is limited to objects fainter than the POSS-II plates 
($r \sim 21$), the USS tecnique is still $2.5$ times more efficient ($\epsilon_{USS} = 
0.52$ vs. $\epsilon_{OPT} = 0.19$).
For an optical search to reach a comparable efficiency it is necessary to 
select objects fainter than $r = 23$, but this implies that about half of the HzRGs 
are lost because of the imposed magnitude bias. 
The advantage of the USS tecnique is that a $\sim 0.5$ search efficiency is 
already reached at the POSS-II plates limit, where all the optical identification work 
is done without telescopes. 
However, this tecnique has the drawback that up to $40\%$ of the 
HzRGs of the sample are lost simply because of the applied spectral index bias. 
Interestingly, the introduction of a strong angular-size bias such as $\theta < 15''$ 
can double the searching efficiency irrespectively of the adopted tecnique, but only 
in the case that no optical bias has been introduced first.
\end{abstract}

\end{frontmatter}

\section{Introduction}

The efforts made in the last three decades to build large samples of Ultra Steep 
Spectrum (USS) radio sources have been justified by a supposed high efficiency 
of this tecnique to select very distant galaxies. 
Its effectiveness derives from a combination of the (high) redshift of the source 
with an intrinsic steepening of the radio spectrum at high frequencies. 
In practice, distant sources with concave radio spectra, will have steeper spectral 
indices than similar sources at low redshift. 
It has been demonstrated (Tielens et al. 1979, Blumenthal \& Miley 1979) that the 
introduction of  a radio spectrum steepness criterium implied a large decrease of 
the fraction of object with optical counterparts identified on the POSS-I ($R \leq 
20$). 
This was consistent with the steeper spectrum sources being, on average, at high 
redshifts. Surveys at low frequencies ($\nu < 400$MHz) are particularly efficient in 
selecting radiogalaxies and have at least two benefits. 
One is that quasars contribute by no more than $20\%$ to the total number counts
; the second is that most of the detected energy arises from the radio-lobes 
with negligible contribution from the sources cores, where Doppler boosting can play 
an important role. 
In principle, as differential sources counts at $400$MHz show an excess over the 
Euclidean prediction in the range $\sim 0.25 - 1$Jy, a survey with such 
limiting fluxes should provide an unbiased sample of objects and a high fraction of 
high-redshift radiogalaxies (henceforth HzRGs) as well. 
The difficulty in picking-up these object arises from the large number of 
candidates to be targeted at optical; given the low intrinsic 
fraction of high-z objects in a radio survey, the majority of the targets will 
reveal to be unwanted foreground objects.
It is thus of fundamental importance to introduce some kind of bias to reduce 
the number of HzRGs candidates.

\noindent
In low-frequency radio surveys ($178-408$MHz), the extragalactic 
sources with steep spectra ($\alpha < -1$) are a small fraction of the 
total, typically no more than $3 \div 5$\%, and this fraction lowers down to $\sim 
0.5$\% if more drastic cuts ($\alpha < -1.3$) are adopted.
Starting from large unbiased radio catalogs is thus essential to build-up conspicuous 
samples of USS sources.
Selecting USS sources implies that the imaging/spectroscopic 
optical follow-up work is limited to a relatively small number of candidates, giving us 
higher chances of success.
In addition, radio searches are not affected by dust/obscurement effects, contrarily 
to what happens with pure optical searches. 
The drawback is that the resulting sample may be not representative of the entire 
population of the high-z objects, since studies conducted on complete samples of 
classical double 
sources (Blundell et al. 1999) showed that sources with steep spectra at low 
frequencies tend to be smaller and more powerful with respect to sources with 'normal' 
spectra.
With the advent of the new all-sky surveys (WENSS, NVSS, FIRST), unprecedently 
deep and large USS samples have been selected (De Breuck et al. 2000), either to 
increase the number stastistics and to detect intrinsically fainter 
sources. 
These surveys virtually should have pushed the steep spectrum tecnique to 
its limit, in the sense that they are deep enough for any AGN class object 
to be detected even at the highest redshifts ($z \sim 7 \div 10$).
Nevertheless, despite much has been published about the radio surveys and 
the properties of the USS sources, very few claims do exist about the 
real effectiveness of this tecnique in selecting HzRGs.
The temptative estimate given by some authors (van Breugel et al. 1997), though  
quite realistic, resulted from 
such a mix of radio/optical/NIR biases that makes it impossible to disentangle the 
role of the radio spectrum steepness itself. The organization of the paper is as 
follows: the properties of the existing samples of USS 
sources are resumed in Sect. 2 and the procedure used to derive 
a quantitative estimate of the real effectiveness of the USS tecnique is described 
in Sect.3. Conclusions are summarized in Sect.4.

\section{USS samples from the literature; radio and optical properties}

Up to now many USS samples have been selected, as many groups 
started extensive searches for very distant radiogalaxies 
using the radio-spectrum steepness as a selection criterium.
In practice, all the existing radio catalogs in the range of frequencies 
$38 - 5000$MHz have been combined together to build USS sources samples. 
When building an USS sample, it is common practice to perform a positional 
cross-correlation of the sources detected in a low-frequency 
survey with those of a high-frequency one, in order to calculate a two-point 
spectral index. 
The selection frequency plays an important role in determining the 
efficiency of finding HzRGs. 
An outstanding example is the sample of $29$ 4C sources (Baldwin \& Scott 1973), whose 
counterparts were 
searched on the $38$MHz survey of Williams et al. (1966) by imposing the criterium 
$\alpha_{38}^{178} < -1.2$.
Twelve out of $29$ sources were found to be associated with Abell clusters 
with median redshift $z = 0.095$.
These low-frequency USS sources are almost exclusively found in rich clusters, and are 
often associated with extended radio haloes or relic structures. Syncrotron losses of an 
ageing population of emitting electrons, confined by the dense intra-cluster medium, 
may explain the unusual steepeness of their spectrum.
On the other hand, radio spectra of the high-z USS sources do steepen at high 
frequencies due to a combination of syncrotron/inverse compton losses and the redshift.
At low frequencies, a flattening of their spectrum may result from syncrotron-self 
absorption occurring in the sources hot-spots.

%
\begin{table*} 
\begin{center}
\caption[]{The most important USS samples actually found in the literature. 
$^{a}$ The Leiden compendium is composed of 9 subsamples. 
REFERENCES: Baldwin-Scott (4C-BS), Baldwin \& Scott (1973); 
Tielens (4C-T), Tielens et al. (1979); 
6C$\ast$, Blundell et al. (1998); 
Ratan-Texas (RC), Parijskij et al. (1991); 
Westerbork-VLA (WV), Wieringa \& Katgert (1992); 
Bologna (B3.2), Pedani \& Grueff (1999); 
Molonglo (MRC), McCarthy et al. (1990, 1991); 
Texas-NVSS, WENSS-NVSS, Molonglo-Parkes (TN, WN, MP), De Breuck et al. (2000); 
Leiden Compendium , R\"{o}ttgering et al. (1994).}
\begin{tabular}{cccccc}
       &            &         &            &                      \\
\hline
Sample & Limit Flux (Jy)& $\alpha$ bias & $\theta$ bias & Sources  \\
\hline
4C-BS &$S_{178} > 2$; $S_{38} > 14$ & $\alpha_{38}^{178} < -1.2$ & none & $29$   \\
       &            &         &            &                     \\
4C-T & $S_{178} > 2$ &  $\alpha_{178}^{1415} < -1.0$; $\alpha_{1415}^{4995} < -1.0$
& none & $34$   \\
       &            &         &            &                     \\
6C$\ast$ & $0.96 < S_{151} < 2.0$ & $\alpha_{151}^{4850} < -0.981$ & $\theta < 15''$ & 
$29$  \\
       &            &         &            &                       \\
RC     & $S_{365} > 0.25$; $S_{3900} > 0.004$ & $\alpha_{365}^{3900} \leq -0.9$ & 
none & $91$   \\
       &            &         &            &                      \\
WV     & $S_{327} > 0.030$ & $\alpha_{327}^{608} < -1.1$ & $\theta < 25''$ & $81$   
\\
       &            &         &            &                      \\
B3.2   & $S_{408} > 0.1$ & $\alpha_{408}^{1400} < -1.0$ & $\theta < 45''$ & $185$  \\
       &            &         &            &                    \\
MRC    & $S_{408} > 0.95$ & $\alpha_{408}^{843} < -0.9$ & none & $150$    \\
       &            &         &            &                            \\
TN     & $S_{365} > 0.15$; $S_{1400} > 0.01$ & $\alpha_{365}^{1400} < -1.3$ & none & 
$268$   \\
       &            &         &            &                      \\
WN     & $S_{325} > 0.018$; $S_{1400} > 0.01$ & $\alpha_{325}^{1400} < -1.3$ & none & 
$343$     \\
       &            &         &            &                     \\
MP     & $S_{408} > 0.70$; $S_{4800} > 0.035$ & $\alpha_{408}^{4800} \leq -1.2$   & none & 
$58$   \\
       &            &               &            &                       \\
Leiden & Various$^{a}$  & $\alpha < -1$ & none   & $605$   \cr
\hline
\end{tabular}
\end{center}
\end{table*}

\noindent
Though most of the USS sources samples from the literature have almost 
complete radio data, their optical follow-up programmes are often far from 
being completed. 
Deriving statistical results from these samples is not straightforward, as 
different selection criteria (e.g. spectral index, sources LAS, optical magnitude) 
were often mixed together to maximize the searching efficiency but the net effect on 
the final samples is not well understood. 
Moreover, the biases introduced by these criteria are almost impossible to 
disentangle as we usually deal with small samples spanning small ranges in radio 
luminosity. This is the reason why very few quantitative claims do exist 
on the real effectiveness of the USS tecnique in selecting high-z galaxies up to date.
In Tables 1 and 2, we summarize radio and optical data of the most 
representative USS sources samples available from the literature.

%
\begin{table*} 
\begin{center}
\caption[]{Radio and optical properties of the USS sources samples listed in Table 1. 
The r magnitudes of the MRC sources were transformed as $R = r - 0.4$}
\begin{tabular}{ccccccc}
       &            &         &            &                  &      \\
\hline
Sample & \%ID on POSS-I & $R_{med}$ & $\theta_{med}$(arcsec) & \%$z_{spec}$ & 
$z_{median}$ & \%$z > 2$ \\
\hline
4C-BS  & $50$       & $n.a.$  & $n.a.$    & $58.0$  & $0.095$ & $0.0$   \\
       &            &         &           &         &        &        \\
4C-T   & $7$        & $22.3$  &  $20.0$   &  $50.0$ & $2.27$ & $53$   \\
       &            &         &           &         &        &        \\
6C*    & $6.0$      & $22.5$  & $4.9$     & $100.0$ & $1.92$ & $41$   \\
       &            &         &           &         &        &       \\
RC     & $15$       & $22.1$  & $12.4$    & $n.a.$  & $n.a.$ & $n.a.$ \\
       &            &         &           &         &        &       \\
WK     & $9.0$      & $22.9$  & $6.0$     & $n.a.$  & $n.a.$ & $n.a.$ \\
       &            &         &           &         &        &       \\
B3.2   & $13.0$     & $24.0$  & $8.1$     & $n.a.$  & $n.a.$ & $n.a.$ \\
       &            &         &           &         &        &       \\
MRC    & $24.1$     & $22.3$  & $9.5$     & $19.3$  & $1.26$ & $30$  \\
       &            &         &           &         &        &       \\
TN     & $13.0$     & $23.2$  & $5.5$     & $7.1$   & $2.42$ & $73$  \\
       &            &         &           &         &        &       \\
WN     & $13.4$     & $22.8$  & $3.5$     & $4.4$   & $2.52$ & $61$  \\
       &            &         &           &         &        &        \\
MP     & $15.5$     & $n.a.$  & $7.8$     & $15.5$  & $1.36$ & $33$   \\
       &            &         &           &         &        &        \\
Leiden & n.a.       & $22.2$  & $10.0$    & $4.6$   & $1.83$ & $46$  \cr
\hline
\end{tabular}
\end{center}
\end{table*}

\noindent
From Table 2 it results that the existing USS samples are very effective 
in selecting very distant galaxies with roughly half of the objects being indeed 
radiogalaxies with redshift $z > 2$.
It should be noted, however, that apart the 6C$\ast$ sample which has complete optical 
information, the above statistics are derived from subsamples which represent only a 
small fraction of the original samples. 
In addition, some 'hidden' selection criteria (e.g. a K-band magnitude lower-limit), 
sometimes adopted to better discard foreground objects (e.g. De Breuck et al 2000), 
make it difficult to assess the real efficiency of such a radio tecnique. 

\noindent
With the only exception of the 3CR, no other unbiased radio 
samples are available with complete radio and optical data as well.
Nevertheless it is interesting to compare the properties of some unbiased samples 
with those of poorly biased samples such as the B3-VLA. 

\begin{table*} 
\begin{center}
\caption[]{Comparison of the HzRGs content of some well studied radio surveys with 
and without radio biases}
\begin{tabular}{cccccccc}
       &       &        &      &       &             &          &     \\
\hline
Survey & Frequency & Flux Density & \# Sources & $z < 2$ & $z \geq 2$ & No 
$z$ & Biases  \\
\hline
3CR    & $178$MHz & $9.0$Jy & $298$ & $99.5\%$ & $0.5\%$ & $0\%$  &  none  \\
       &          &         &       &    &         &        &            \\
MRC    & $408$MHz & $0.95$Jy & $558$ & $93\%$ & $7\%$   & $40\%$ &  none  \\
       &          &          &        &     &         &        &        \\
B2/1Jy & $408$MHz & $1.0 < S < 2.0$Jy & $59$ & $87.7\%$ & $12.3\%$ & $10\%$ 
& none \\
       &          &         &        &          &          &          &      \\
B3-VLA & $408$MHz & $S > 0.8$Jy  & $120$  &  $80\%$  & $20\%$  & $28\%$ & 
$\alpha<-0.8$ \\
       &        &         &        &          &          &      &  $\theta < 20''$ \cr
\hline
\end{tabular}
\end{center}
\end{table*}

From Table 3, it results that the radio steepness bias, eventually coupled with
an angular-size bias, increases the efficiency of selecting radio galaxies at
redshifts $z > 2$. Excluding the 3CR that contains only one RG with $z >
2$, the other low-frequency radio surveys contain typically no more than
$\sim 10\%$ of HzRGs. 
One of the most studied samples is the B2/1Jy 
(Allington-Smith 1982) which contains $59$ radio sources with $1
\leq S_{408} \leq 2$Jy. Out of $50$ radiogalaxies, $42$ have
redshift information. This subsample has a median redshift
$z = 0.798_{-0.21}^{+0.22}$ and median spectral index $\alpha_{408}^{1400} =
-0.73_{-0.02}^{+0.05}$. 
It contains only $2$ USS sources with $\alpha_{408}^{1400} < -1$, 
of which one is a radiogalaxy with $z = 0.79$ and the other is unidentified. 
This sample includes $12$ RGs with $1 \leq z <2$, 
five with $2 \leq z < 3$ and one with $z > 3$; thus 
$6$ out of $42$ ($14\%$) B2/1Jy sources are indeed high redshift 
galaxies, though none of them would have been selected on the basis 
of the radio spectral index criterium. 
In the case of the B2/1Jy sample, the radio steepness criterium clearly fails to
select the high-z radiogalaxies but, interestingly, if an angular-size cut like 
$\theta < 15''$ is introduced, a total of $30$ objects are selected, including $6$ RGs 
with $z > 2$. 
Thus, the angular-size bias itself would select all the B2/1Jy HzRGs with a 
global efficiency of $0.2$.

\noindent
The B3-VLA sample (Vigotti et al. 2003; in preparation) has a mild spectral index bias, 
coupled with a relatively strong angular-size bias; for this reason it may 
be considered half way between the usual USS samples and the unbiased samples. 
For a typical low-frequency radio sample, a spectral index cut like $\alpha < -0.8$ 
roughly implies half of the sources to be filtered out, while an angular-size 
bias such as $\theta < 20''$ only implies $\sim 20\%$ of the sources to be excluded 
(see also Table 1).
Since low-frequency radio samples all have small median sizes 
($\sim 10''$), 
we can argue that the angular-size bias adopted for the B3-VLA sample has little 
influence on the final content of HzRGs. 
This result suggests that the expected increase of the fraction of $z > 2$ galaxies 
is not seen in the B3-VLA sample mainly because of the mild spectral index cut adopted.

\section{Testing the effectiveness of the USS tecnique}

At present there are no radio surveys with complete radio and optical data that can be 
used to assess the validity of the USS searching technique (see Tables 2 and 3). 
The ideal sample to test the real effectiveness of this tecnique  
would be that with complete radio and optical information. 
Unfortunately, apart the well known 3CR and B2/1Jy samples, no other survey 
has a follow-up programme that reached such a degree of completion. 
While for the 3CR sample the very high limiting flux density virtually prevents any 
HzRG to be selected, in the case of the B2/1Jy sample the poor number statistics 
prevents any statistically significant result to be derived. 

\subsection{The MRC/1Jy sample - radio and optical data}

Actually, the only available low-frequency sample, free of selection biases 
and with a reasonably complete radio/optical dataset is the MRC/1Jy sample
selected by McCarthy et al.(1996) from the Molonglo Reference Catalogue (Large et al. 
1981).
In the following it will be shown that, despite a substantial 
redshift incompleteness toward the faintest magnitudes, this sample 
can provide a dataset five times larger than the B2/1Jy sample, well suited for 
statistical studies. VLA maps at $6$cm with typical resolution of 
$1 - 3''$ exist for all the MRC galaxies which allow a precise morphological 
classification. 
Kapahi et al. (1998) listed the flux densities at $408$MHz and $4.86$GHz for $446$ radio 
galaxies drawn from the MRC/1Jy sample.
When the flux density at $4.86$GHz was not available, the value at $1.4$GHz 
was reported. 
To build-up a homogenous sample, essential for the 
present study, it was decided to get NVSS (Condon et al. 1998) 
flux densities for all the MRC 
radiogalaxies in order to re-define the spectral indices as $\alpha_{408}^{1400}$.
This criterium 
closely resembles the original one adopted to select the MRC USS sample 
($\alpha_{408}^{843}$, see Table 1). 
As the median spectral index for the $446$ MRC galaxies resulted 
to be $\alpha_{408}^{1400} = -0.85 \pm 0.01$, a 'canonical' spectral index bias 
$\alpha_{408}^{1400} < -1$ has been chosen to select this newly defined 
MRC USS sample. 
McCarthy et al. (1996) reported a $96\%$ identification fraction of the MRC radio 
sources down to the $r = 25$ magnitude. 
To date, r-band magnitudes are available for $334$ of these $446$ MRC 
galaxies and redshift information is available for $225$ of 
these $334$ galaxies (McCarthy et al. 1996, Kapahi et al. 1998 and references 
therein). 
Thus, complete radio and optical data actually exist for a subsample of $225$ MRC 
radiogalaxies. Hereinafter the two samples of $334$ and $225$ RGs will be 
referred to as MRCB and MRCC respectively.

\begin{figure}[here]
\includegraphics*[scale=0.45,angle=-90]{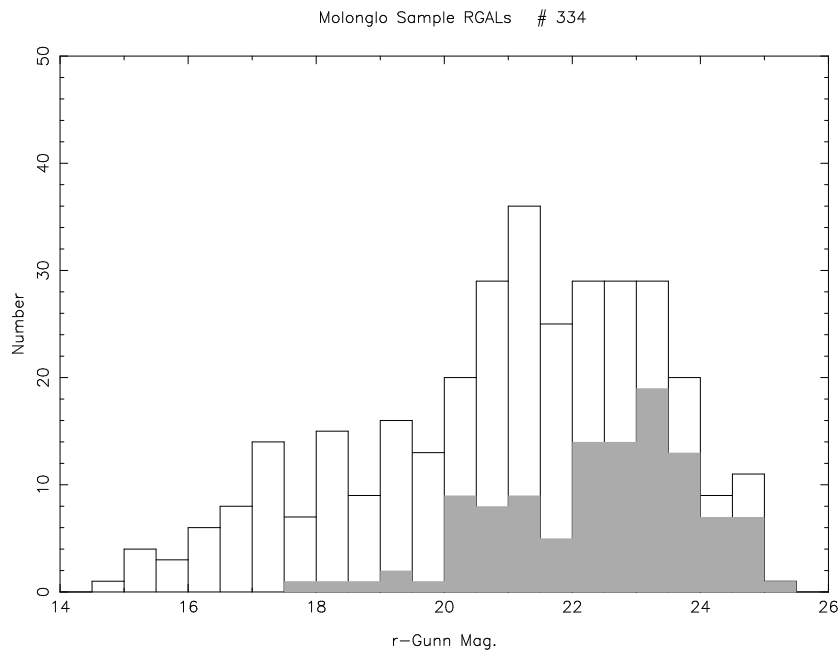}
\caption{The r-band magnitudes distribution for MRCB radio galaxies. 
The shaded area indicates the $109$ RGs for which no redshift information is available}
\end{figure}
\vskip 10pt

\noindent
The {\it r}-band magnitude distribution for the MRCB sample has a median value 
$r = 21.3 \pm 0.1$ and it is shown in Figure 1. 
For comparison the median magnitude of the MRCC sample is $r = 20.7 \pm 0.2$. 
A redshift incompleteness toward the faint end is evident from Figure 1 as 
the $109$ sources lacking of redshift data 
preferentially lie close to the MRC optical identification limit, as confirmed by 
their significantly higher-than-average median magnitude ($r = 22.7_{-0.2}^{+0.1}$). 

\noindent
The redshift incompleteness of the MRCB sample can be assessed as follows. 
After dividing the sample into two subsamples, 
respectively fainter and brighter than the median magnitude, for each subsample the 
ratio between the number of RGs with and without redshift is calculated.
In the bright sample, only $16\%$ of the RGs have no measured redshift 
but the fraction increases to $51\%$ for the faint sample, indicating a 
substantial incompleteness. 
The MRCC sample redshift distribution is shown in Figure 2. 
The median redshift is $z = 0.600_{-0.017}^{+0.034}$, but the objects 
fainter than the median magnitude $r = 20.7$, have a significantly higher 
redshift $z = 1.03_{-0.1}^{+0.09}$.

\begin{center}
\begin{figure}
\includegraphics*[scale=0.45,angle=-90]{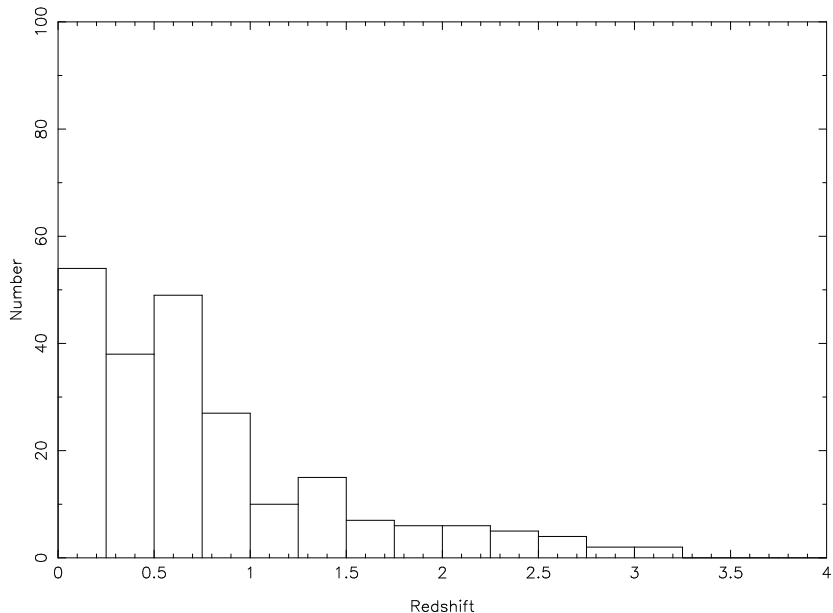}
\caption{Redshift distribution of the MRCC sample}
\end{figure}
\end{center}

A crucial point for our analysis is that the redshift 
incompleteness of the MRCB sample at faint levels could imply 
the MRCC sample has a lower-than-average content of USS sources. 
The fact that the median magnitude of the $109$ RGs with no redshift 
data is similar to that of the known USS samples ($R \sim 23$) 
could favour the above hypothesis. 
Nevertheless it can be demonstrated that this is not the case. 
According to our criterium, $24$ out of the $109$ RGs are also USS 
sources ($22\%$). 
A direct comparison between the USS content of the MRCB and the 
smaller MRCC sample gives respectively the values of $18\%$ and $16\%$; 
thus, the redshift incompletenss of the MRCB sample is unlikely to affect 
the MRCC sample, since only $2\%$ of USS sources are lost.

\subsection{HzRGs and USS sources in the MRCC sample}

The MRCC sample contains $19$ RGs with redshift $z \geq 2$ ($8.4\%$ of the
total), $12$ of which are USS sources. As the sample contains $36$ USS sources, 
if no other selection criterium than the spectral index 
bias is applied, the efficiency would be $0.33$. This relatively high 
efficiency of the USS tecnique has the drawback that a significant fraction of 
genuine targets are lost; in this case, $7$ out of $19$ HzRGs ($\sim 37\%$).

\subsection{The optical magnitude - radio spectral index plane}

As the digitized POSS-I or POSS-II plates are normally used for the first step of 
any optical identification programme, let us assume that all the optical 
counterparts have been identified down to $R \sim 20.5$. 
In the case of the MRCC sample, this implies a reduction of 
the number of HzRGs candidates by a factor of $1.7$, with a consequent increase 
of the searching efficiency 
(now $\epsilon_{OPT} = 0.14$ instead of $\epsilon_{OPT} = 0.08$).
At the same time, the efficiency of the USS technique also increases, reaching the 
considerable value of $\epsilon_{USS} = 0.48$.
Thus, at least in the case of the MRCC sample, we conclude that by simply selecting 
the USS sources fainter than $R \sim 20.5$ one is guaranteed that half of the 
candidates will be HzRGs. 
To study the effect of different radio and optical biases on the MRCC sample, 
the radio spectral indices $\alpha_{408}^{1400}$  were plotted 
against the {\it r}-band magnitudes (see Figure 3). 
For comparison, data for the $109$ MRC galaxies without redshift are also plotted.

\begin{figure}
\includegraphics*[scale=0.45,angle=-90]{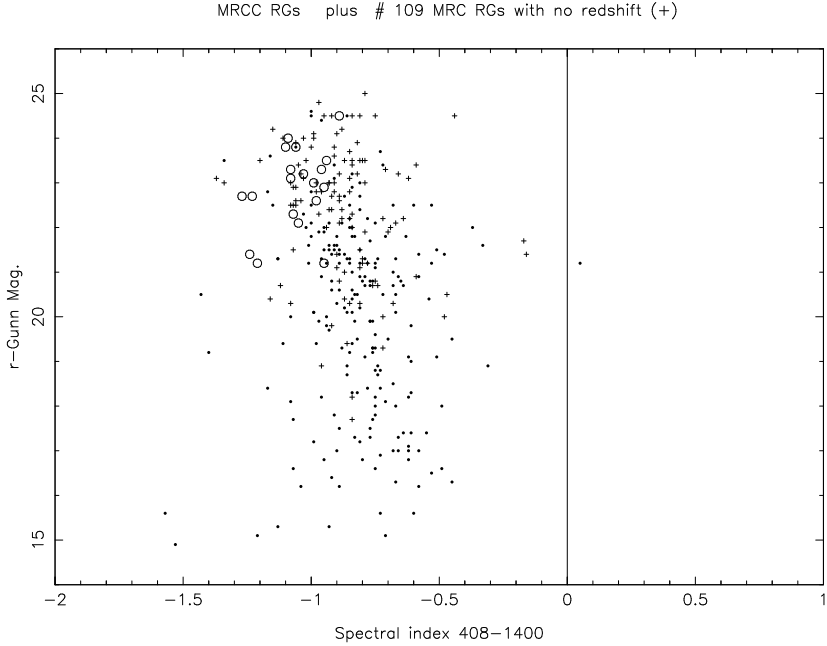}
\caption{The {\it r}-band magnitude against the spectral index $\alpha_{408}^{1400}$ 
for the $225$ MRCC RGs (filled circles); the MRCC HzRGs are indicated with 
open circles and the $109$ MRC RGs without redshift are indicated with crosses}
\end{figure}

\noindent
Since all the MRCC HzRGs are fainter than $r = 21$, 
this value may be taken as a good starting point for our analysis. 
If the RGs with $r \geq 21$ are considered, a total of $101$ objects are 
selected; their spectral indices distribution is shown in Figure 4.

\begin{figure}
\includegraphics*[scale=0.45,angle=-90]{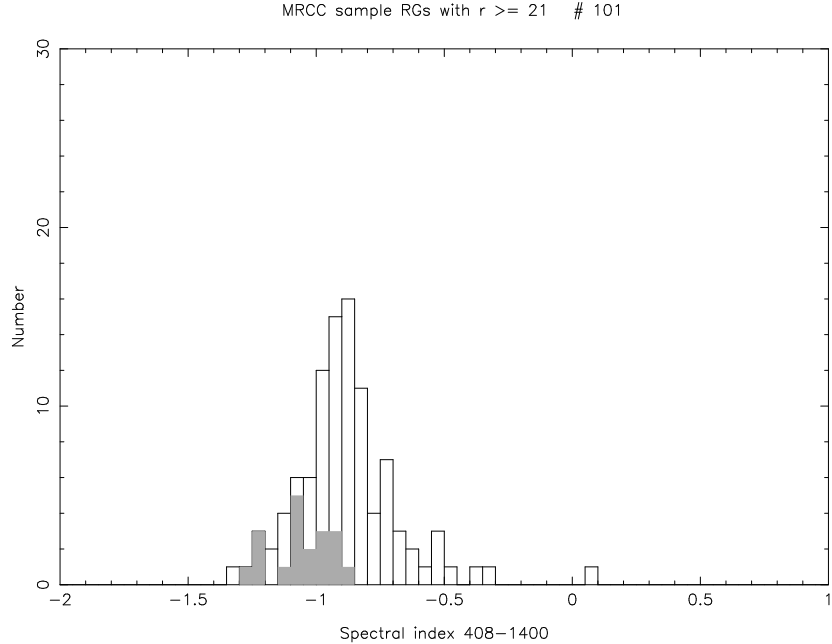}
\caption{Spectral indices distribution of the $101$ MRCC RGs with 
$r \geq 21$; the $19$ MRCC HzRGs are also shown (filled area)}
\end{figure}

If only the USS sources would be targeted at optical, 
$12$ out of $23$ objects would result to be HzRGs, with a net searching efficiency 
$\epsilon_{USS} = 0.52$. Note however that $7$ HzRGs ($37\%$ 
of the total) would be lost as they are not USS sources.
In contrast, if no radio spectrum criterium is adopted, there would be $101$ 
sources to observe at optical ($4$ times more telescope time); 
all the $19$ HzRGs would be selected, resulting in a final selection 
efficiency $\epsilon_{OPT} = 0.19$. 
Thus, when the search is limited to targets fainter than $r = 21$ magnitude, 
the USS searching technique is about $2.5$ times more efficient with respect to 
the optical search. As expected, the overall telescope time investment 
is largely in favour of the radio selection technique.
To increase the efficiency of the optical search, the 
number of candidates to be imaged could be reduced by adopting a more drastic bias 
in magnitude.Lowering the optical threshold down to $r = 22$ 
reduces the sample to $53$ objects (see Figure 5).

\begin{center}
\begin{figure}
\includegraphics*[scale=0.45,angle=-90]{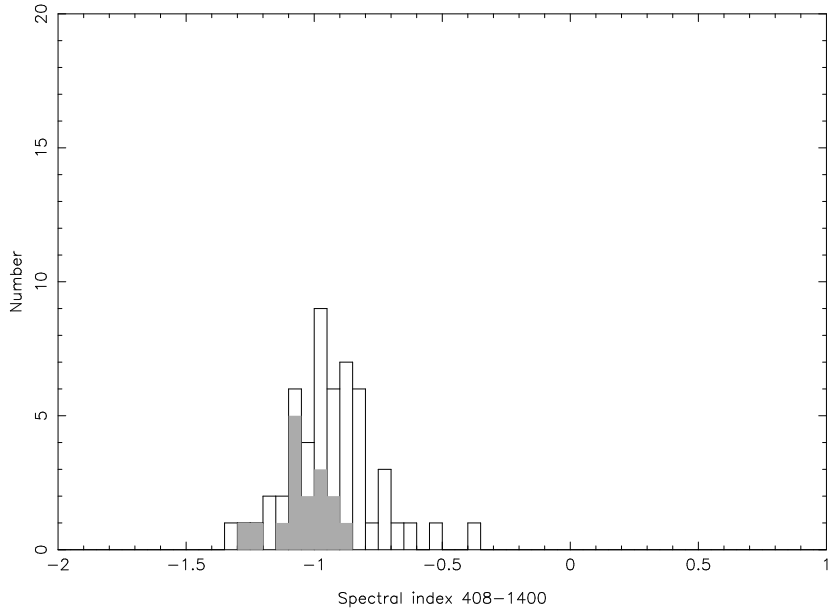}
\caption{Spectral index distribution of the $53$ MRCC RGs with $r \geq 22$. 
The $16$ MRCC HzRGs are also shown (filled area)}
\end{figure}
\end{center}

\noindent
With respect to the original MRCC sample, the number of objects is now 
reduced by a factor of four. Observing all these objects at optical ensures 
that $16$ out of the $19$ HzRGs contained in the MRCC sample 
are selected, with a net efficiency $\epsilon_{OPT} = 0.30$ 
($3$ HzRGs are lost because of the magnitude limit). 
On the other hand, this subsample contains only $17$ USS sources, out of which $10$ 
are HzRGs. The USS technique thus has a remarkable efficiency 
$\epsilon_{USS} = 0.59$ but $6$ HzRGs ($37\%$ of the total) are lost 
as they are not USS sources. 

Note that while $\epsilon_{OPT}$ tends to increase when going to 
fainter magnitudes, $\epsilon_{USS}$ remains almost constant around 
the value $\epsilon_{USS} \sim 0.55$. 
Pushing the optical magnitude limit toward fainter values is not recommended as 
the size of the resulting sample (and its statistical strenght) drastically reduces 
and the fraction of HzRGs that are lost rapidly increases. 
For example, going to $r \geq 23$ implies that only $23$ RGs are retained. 
The optical search reaches its peak efficiency $\epsilon_{OPT} = 0.43$ 
with $10$ HzRGs being selected. 
However $9$ out of $19$ HzRGs ($\sim 50\%$) are lost as they are brighter than 
$r = 23$. At the same time, the USS technique selects $6$ HzRGs from a total 
of $11$ sources, with an efficiency 
$\epsilon_{USS} = 0.54$. In this case $13$ out of $19$ HzRGs
($\sim 70\%$ of the total) are lost. 
Despite the expected loss of some HzRGs, the USS technique is again
more efficient with respect to the optical search by about $25$ percent.

\subsection{The role of the angular-size bias}

Most of the existing USS samples were derived without imposing any angular-size bias 
(see Table 1). 
Nevertheless, even choosing unresolved sources in low-frequencies surveys 
such as Texas or WENSS, automatically implies an intrinsic angular cut-off of 
$\sim 2^{\prime}$ and $\sim 1^{\prime}$ respectively. 
The samples completeness may be affected by other factors. 
In the case of the Texas survey, the complicate behaviour of the beam 
of the interferometer (Douglas et al. 1996) implies that the resulting TN sample 
is only $40\%$ complete down to the survey nominal flux density limit 
(De Breuck et al. 2000).

\begin{figure}
\includegraphics*[scale=0.5,angle=-90]{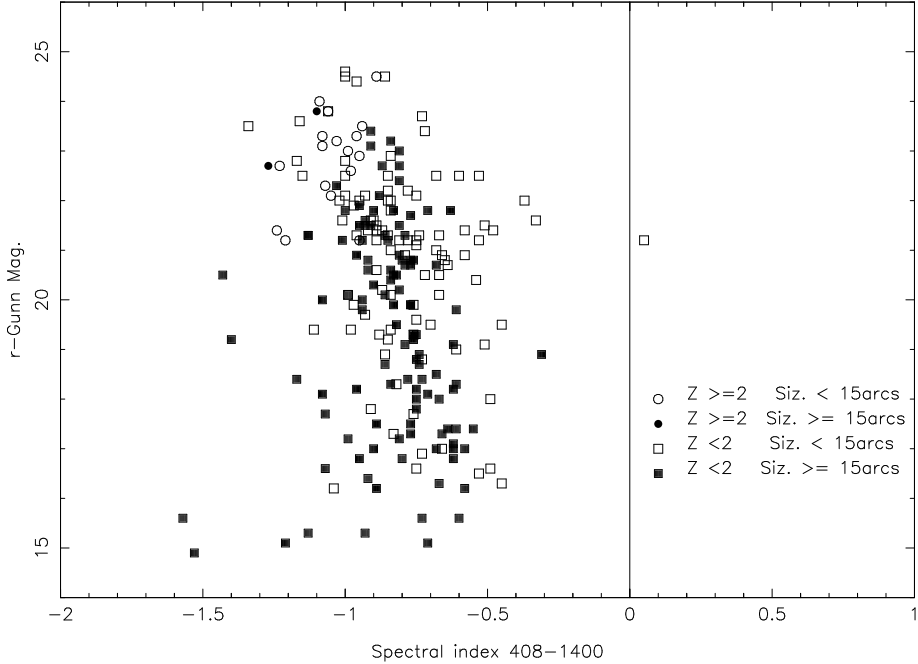}
\caption{Spectral indices distribution of the $225$ MRCC RGs as a 
function of the {\it r}-band magnitude; filled symbols indicate objects with 
$\theta \geq 15''$. The $19$ MRCC HzRGs are indicated with circles}
\end{figure}

To increase the efficiency of the USS tecnique some authors introduced a 
angular-size criterium to filter-out most of the foreground objects. 
Since the most distant objects should be among the faintest and the smallest ones, 
an angular size bias should increase the efficiency of finding HzRGs. 
The choice of the optimal cut-off value is quite arbitrary; though there are very few 
cases of known HzRGs with large angular sizes (\astrobj{4C 23.56} at $z = 2.483$ has 
$\theta 
= 53''$), most of them have $\theta < 35''$ (Carilli et al. 1997). 
Adopting more drastic criteria such as $\theta < 15''$, implies a reduction 
by $\sim 30\%$ of the total number of sources but it can lead to the exclusion 
of several good candidates. 
It can be demonstrated that the angular-size bias alone is effective to filter-out 
relatively bright and thus nearby objects. 
In fact, if all the MRCC objects are considered without any magnitude 
and size limit (see Fig.6), the optical search efficiency is only 
$\epsilon_{OPT} = 0.08$, while the USS efficiency is $\epsilon_{USS} = 0.33$, four 
times higher. 
If a $\theta < 15''$ bias is introduced, we found that $\epsilon_{OPT}$ 
and $\epsilon_{USS}$ become almost twice as large (respectively 
$0.15$ and $0.53$). 

\begin{table*}
\begin{center}
\caption[]{Comparison of the relative efficiencies $\epsilon_{OPT}$ and
$\epsilon_{USS}$ (as derived from the MRCC sample) in selecting HzRGs as a 
function of different values of the limiting magnitude. 
Given the small number of objects at $r \geq 23$, little significance should be 
attributed to this case. The effect of 
the angular-size bias is shown for the case in which no magnitude limit is
imposed to the sample.}
\begin{tabular}{ccccccc}
       &       &    &     &           &             &    \\
\hline
       &       &    &      &          &             &    \\
Mag. Bias & Size Bias  & \# Sources & $\epsilon_{OPT}$ & HzRGs lost
& $\epsilon_{USS}$ & HzRGs lost  \\
       &      &         &        &    &             &                \\
 none & none  & $225$   & $0.08$ & none & $0.33$ & $0.37$             \\
       &      &         &   &      &             &                \\
 none &$\theta < 15''$ & $112$ & $0.15$ & $0.10$ & $0.53$ & $0.17$ \\
       &      &         &   &      &             &                \\
$r \geq 20$ & none & $135$ & $0.14$ & none & $0.48$ & $0.37$  \\
       &      &         &   &      &             &                \\
$r \geq 21$  & none  & $101$ & $0.19$ & none & $0.52$  & $0.37$ \\
       &      &   &      &         &            &                \\
$r \geq 22$  &none & $53$  & $0.30$ & $0.16$ & $0.59$ & $0.37$ \\
       &      &  &       &         &            &                \\
$r \geq 23$  &none & $23$  & $0.43$ & $0.50$ & $0.54$ & $0.70$ \\
       &    &   &  &         &             &                \cr
\hline
\end{tabular}
\end{center}
\end{table*}

On the other hand, the angular-size bias seems to have little or no effect when 
an optical bias has been formerly introduced. 
Let us consider the previously discussed case when the MRCC objects 
fainter than $r = 22$ were selected. 
Again, if the above angular-size bias is applied, 
the efficiency of the optical and USS searches 
remains almost unchanged, being respectively $\epsilon_{OPT} = 0.31$ and 
$\epsilon_{USS} = 0.57$. 
This is somewhat expected since an optical bias like $r > 22.0$ already selects 
those objects wich are, in principle, more distant and thus smaller. 

\noindent
To conclude, the introduction of a strong angular size bias such as 
$\theta < 15''$, increases 
the efficiency in selecting HzRGs by a factor of two, independently 
on what searching technique is adopted (optical or radio). 
Note that the same increase of efficiency is obtained by selecting only 
those objects with $r \geq 20$, irrespectively of their angular size. The relative 
efficiencies of the optical and radio searching techniques as a function 
of different cuts in the magnitude and angular size distribution are 
summarized in Table 4.

\section{Conclusions}

In this paper the real efficiency of the Ultra Steep Spectrum tecnique 
in finding HzRGs is quantitatively demonstrated for the first time since 
this radio tecnique was introduced. 
Moreover, an updated view is given of the radio/optical properties 
of the most representative USS sources samples and the role of the spectral index and 
angular-size biases is discussed. 
A new sample of USS sources was selected from the MRC/1Jy sample (McCarthy et al. 1996; 
Kapahi et al. 1998). Since the available radio dataset is not homogeneous, 
the radio flux densities from the NVSS were used to re-calculate the spectral 
indices as $\alpha_{408}^{1400}$ and a conservative radio-spectrum bias 
$\alpha_{408}^{1400} < -1.0$ was adopted to select the USS sources. 
Our USS sample thus consists of $225$ MRC radiogalaxies with complete optical 
and radio data and it was used as a benchmark to study how different radio and optical 
biases do affect a homogeneous population of radio galaxies. 
It was found that the USS searching technique 
is much more efficient in finding HzRGs than a conventional search based on 
the optical magnitudes of the RGs. 
If no restrictions are applied to the optical magnitudes distribution, the 
USS technique may be up to $4$ times more efficient than the optical one. 
The obvious drawback is that a substantial fraction of the HzRGs population 
may be lost because of the introduced radio spectrum bias. 
When considering galaxies fainter than magnitude $r = 20$, the radio-steepness criterium 
guarantees that $\sim 50\%$ of the candidates are indeed HzRGs; this has to be compared 
with a mere $\sim 15\%$ of the optical search. 
Going to fainter magnitudes leads to a progressive increase of the efficiency 
of the optical search, that reaches its peak at $r \geq 23$ with a notable 
efficiency $\epsilon_{opt} = 0.43$ (see Table 4).
Note, however, that half of the HzRGs of the sample are lost when applying such a 
strong optical bias. 
Interestingly, the efficiency  of the USS tecnique $\epsilon_{USS}$ remains stable 
around $0.50 \div 0.55$ irrespectively of the applied optical magnitude bias; 
going to fainter magnitudes only increases the fraction of 
genuine HzRGs that are lost (up to $70\%$ at $r \geq 23$).
To conclude, even at the faintest optical magnitudes here studied, the USS tecnique is 
$25\%$ more efficient of an optical search in selecting HzRGs, with the remarkable 
difference that the radio tecnique already reaches a $\sim 0.5$ efficiency at 
the $r \sim 20$ magnitude limit. 
The advantage of the optical search is that, at least up to $r = 22$, a much smaller 
fraction of HzRGs is lost with respect to the radio tecnique 
($\sim 15\%$ instead of $\sim 40\%$). 
The introduction of an angular-size bias to the MRCC sample, such as $\theta < 
15''$, increases the searching efficiency by a factor of two, irrespectively of the 
searching tecnique adopted. 
However this additional bias has little or no effect when a magnitude bias such 
as $r > 20$ is applied first to the sample, as it is the case when the first step of 
the optical identification program is done on the digitized POSS-I or POSS-II plates.
This evidence suggests that the angular-size bias, sometimes used as an additional 
criterium, plays a secondary role with respect to the spectral index bias. 
This is also suggested by the finding that 
the USS samples actually adopting the steepest spectral index criterium 
and no angular-size bias (TN and WN samples; $\alpha < -1.3$) are those with the 
highest efficiencies in finding HzRGs.

\section{Acknoledgements}
The author wants to thank the anonymous referee and the NA editor, Prof. G. Setti, for 
their useful comments and suggestions on the manuscript.

\end{document}